\documentclass[a4paper]{article}
\usepackage{ISCSLP2024}
\usepackage{ifthen}
\newboolean{blind}
\usepackage{url}
\usepackage{hyperref}
\usepackage{makecell}

\setboolean{blind}{false} 
\title{The ISCSLP 2024 Conversational Voice Clone (CoVoC) Challenge: Tasks, Results and Findings}
\name{
	\ifthenelse{\boolean{blind}}{Anonymous to ISCSLP}
	{Kangxiang Xia$^1$,Dake Guo$^1$, Jixun Yao$^1$, Liumeng Xue$^2$, Hanzhao Li$^1$, Shuai Wang$^3$, Zhao Guo$^1$, Lei Xie$^1$,Qingqing Zhang$^4$,Lei Luo$^4$,Minghui Dong$^5$,Peng Sun$^6$}
}

\address{
  \ifthenelse{\boolean{blind}}{Anonymous to ISCSLP}
  {
  	$^1$Audio, Speech and Language Processing Group (ASLP@NPU), School of Computer Science, \\ Northwestern Polytechnical University, Xi'an, China\\
    $^2$School of Data Science, The Chinese University of Hong Kong,\\
    Shenzhen(CUHK-Shenzhen), China \\
    $^3$Shenzhen Research Institute of Big Data (SRIBD) $^4$Magic data \\
    $^5$Insitutue for Infocomm Research (I2R), Singapore $^6$China Computer Federation 
  }
}

\email{
	\ifthenelse{\boolean{blind}}{Anonymous to ISCSLP}
	{xkx@mail.nwpu.edu, coauthor@company.com}
}

\begin{document}

\maketitle
\begin{abstract}
The ISCSLP 2024 Conversational Voice Clone (CoVoC) Challenge aims to benchmark and advance zero-shot spontaneous style voice cloning, particularly focusing on generating spontaneous behaviors in conversational speech.
The challenge comprises two tracks: an \textit{unconstrained track} without limitation on data and model usage, and a \textit{constrained track} only allowing the use of constrained open-source datasets. A 100-hour high-quality conversational speech dataset is also made available with the challenge.
This paper details the data, tracks, submitted systems, evaluation results, and findings. The challenge's official website is https://www.magicdatatech.com/iscslp-2024.

\end{abstract}

\section{Introduction}


Text-to-speech (TTS) aims to generate speech that sounds as natural and human-like as possible. Recent advancements in neural speech synthesis have significantly enhanced the quality and naturalness of generated speech~\cite{tacotron,fs,vits}, leading to widespread applications of TTS systems in real-world scenarios. A notable breakthrough in the field is witnessed in zero-shot TTS, driven by expanded datasets~\cite{tortoise} and new approaches~\cite{valle} (e.g., decoder-only paradigms), attracting extensive attention from academia and industry. However, these advancements haven't been sufficiently investigated to address challenges in spontaneous~\cite{spontts,sponlm} and conversational~\cite{congjian_conversation} contexts. Specifically, the primary challenge lies in effectively managing prosody details in the generated speech, which is attributed to the diverse and intricate spontaneous behaviors that differentiate spontaneous speech from read speech.

Large-scale TTS systems yield promising outcomes in zero-shot generation due to in-context learning ability. However, a prevalent challenge in the field of large-scale zero-shot TTS is the lack of consistency in training and testing datasets, along with a standardized evaluation benchmark. This issue hinders direct comparisons and makes it challenging to accurately assess various systems' performance.

We launch the Conversational Voice Clone Challenge (CoVoC) to promote the development of expressive spontaneous-style speech synthesis in the zero-shot scenario. 
Besides the existing 10,000-hour WenetSpeech4TTS~\cite{wenetspeech4tts}  dataset and 180 hours of Mandarin conversational speech data~\footnote{\url{https://www.openslr.org/123/}}, we also release a new 100-hour high-quality conversational dataset.
Furthermore, we also conduct a standardized testing dataset accompanied by carefully designed text which aims to establish a comprehensive benchmark. This paper presents the data details, track design, submitted systems, evaluation results, and key findings. 


\section{Challenge Design}
\label{sec:task}


The goal of the CoVoC is to conduct a benchmark for zero-shot voice cloning with conversational speech, aiming to evaluate and compare the performance of different systems in generating conversational speech.
\subsection{Track Setting}
The CoVoC challenge has two tracks: a constrained track and an unconstrained track. In the constrained track, only the specified dataset can be used for model training, while pre-trained open-source models are allowed. The details of training datasets are as follows:

\begin{itemize}
\item \textbf{WenetSpeech4TTS~\cite{wenetspeech4tts}}: A multi-domain corpus derived from the open-sourced WenetSpeech~\cite{wenetspeech} dataset. Tailored for TTS tasks, we refined WenetSpeech by adjusting segment boundaries, enhancing the audio quality, and eliminating speaker mixing within each segment. Following a more accurate transcription and quality-based data filtering process, the obtained corpus contains 12,800 hours of paired audio-text data. 
\item \textbf{MAGICDATA} (Conversational Speech Corpus): A 180-hour speech dataset recorded with various mobile devices. A total of 663 speakers were invited to participate in the recording. Recordings were conducted in a quiet indoor environment. All speech data were manually labeled and professional inspectors proofed the transcriptions to ensure the labeling quality.
\item \textbf{HQ-Conversations}: A 100-hour dataset featuring 200 speakers, including 75 males and 125 females. The content consists of segmented conversations, which closely reflect daily life scenarios, characterized by natural and expressive language rather than a scripted or read-aloud style. The dataset has undergone rigorous screening and verification to ensure high accuracy.
\end{itemize}
The unconstrained track places no limitations on the training process, allowing participants to use any available data and techniques to train their models.

\begin{table*}[ht]
  \caption{Overview of submitted systems in CoVoC. AR and NAR are denoted as autoregressive modeling and non-autoregressive modeling, respectively.}
  \label{tab:total_submint}
  \centering
  \footnotesize
  \begin{tabular}{c c c c c c c c c c}
    \toprule
    \textbf{ ID}& \textbf{Team Name} & \makecell{\textbf{Train-data} \\ \textbf{(hours)}}& \textbf{Acoustic Model} & \textbf{AR/NAR} &  \makecell{\textbf{Waveform} \\ \textbf{(Generation)}}& \textbf{CER ↓} & \textbf{SIM ↑} & \textbf{Final Score ↑} \\
    \midrule
    U0 & Official         & 80k   & TorToise~\cite{tortoise}-like     & AR    & Vocoder & 6.86 & 0.849 &  \textbf{3.91} \\
    \midrule
    U1 &MASTER           & 50k   & Mega-TTS~\cite{megatts}          & NAR   & Vocoder & \textbf{2.56} & 0.843 &  \textbf{3.83} \\
    U2 &C-TTS            & 30k   & VALL-E~\cite{valle}-like       & AR    & Codec   & 5.46 & 0.852 & 3.77 \\
    U3 &Orion            & 10k   & DelightfulTTS~\cite{delightfultts}     & NAR   & Codec  & 3.08 & 0.890 & 3.75 \\
    U4 &Sigma            & ---   & TorToise-like     & AR    & DiTs~\cite{dit}    & 3.89 & 0.808 & 3.75 \\
    U5 &zyzx\_AI          & ---   & GPT-SoVITS        & AR    & SoVITS & 3.99 & 0.848 & 3.72 \\
    U6 &Fish\_Audio  & 300k  & Fish Speech    & AR    & VITS~\cite{vits}   & 7.1 & 0.867 & 3.65\\
    U7 &hySoundClone     & 700   & Bert-VITS2-like   & NAR   & Vocoder & 4.79 & \textbf{0.894} &  3.52\\
    C1 &we\_are\_NPC       & --    & GPT-SoVITS        & AR    & SoVITS & 13.77 & 0.817 & 3.71 \\
    C2 &Fish\_Audio  & --    & Fish Speech       & AR    & VITS & 7.18 & 0.867 & 3.63 \\
    C3 &THU-HCSI         & --    & MusicGen~\cite{musicgen}-like     & AR    & Codec & 10.29 & 0.797 & 3.61 \\
    C4 &ViveTTS          & --    & SPEAR-TTS~\cite{speartts}-like                & AR    & Flow Matching\cite{flowmatching} & 13.36 & 0.841 & 3.27 \\
    C5 &SMIIP\_TTS        & --    & LauraGPT~\cite{lauragpt}          & AR    & Codec & 34.28 & 0.760 & 3.14 \\
    
    \bottomrule
  \end{tabular}
\end{table*}

\subsection{Test Datasets}
We carefully selected 20 speech samples from 20 speakers as the target speaker's prompt during the testing phase.
The target speakers are categorized into two types based on their speaking style: 8 speakers with an ordinary reading style and 12 speakers with a conversational spontaneous style. This setup is intended to see how systems perform on different prompts in zero-shot conversational speech generation. The lengths of the audio prompts ranged from 5 seconds to over 20 seconds, and the distribution of audio lengths is shown in Table~\ref{tab:audio_len}.

\begin{table}[h]
  \caption{Length distribution of the selected audio prompts.}
  \label{tab:audio_len}
  \centering
  \begin{tabular}{ r@{\hspace{1em}}r  r }
    \toprule
    \textbf{Audio Length (S)}    && \textbf{Number of Prompts}    \\
    \midrule
    \multicolumn{2}{c}{0--5}   & \multicolumn{1}{c}{6}  \\
    \multicolumn{2}{c}{5--10}  & \multicolumn{1}{c}{8}  \\
    \multicolumn{2}{c}{10--25} & \multicolumn{1}{c}{6}  \\
    \midrule
    \multicolumn{2}{c}{Total} & \multicolumn{1}{c}{20}  \\
    \bottomrule
  \end{tabular}
\end{table}

The text test sets for the CoVoC challenge include seven distinct types of text: conversational texts, colloquial texts, stammer texts, rhotic accent texts, tone sandhi texts, polyphonic character texts, and long-form audiobook texts. To ensure impartiality and prevent specialized handling, the exact categories of texts are not specified in the test set. The distribution of these text types is detailed in Table~\ref{tab:text_cat_dis}. Each participating team is required to synthesize audio for all test texts across 20 target speakers, resulting in a total of 8,600 synthetic audio samples.

\begin{table}[h]
  \caption{Distribution of category of the test texts.}
  \label{tab:text_cat_dis}
  \centering
  \begin{tabular}{ r@{\hspace{1em}}r  r }
    \toprule
    \multicolumn{2}{c}{\textbf{Text Category}} & \textbf{Number of Texts} \\
    \midrule
    \multicolumn{2}{c}{Conversational (short,middle,long)}   & \multicolumn{1}{c}{270}  \\
    \multicolumn{2}{c}{Colloquial}                & \multicolumn{1}{c}{51}  \\
    \multicolumn{2}{c}{Stammer}      & \multicolumn{1}{c}{24}  \\
    \multicolumn{2}{c}{Rhotic accent}                    & \multicolumn{1}{c}{19}  \\
    \multicolumn{2}{c}{Tone sandhi}                         & \multicolumn{1}{c}{16}  \\
    \multicolumn{2}{c}{Polyphonic Character}                & \multicolumn{1}{c}{44}  \\
    \multicolumn{2}{c}{Long-form Audiobooks}                  & \multicolumn{1}{c}{6}  \\
    \midrule
    \multicolumn{2}{c}{Total}                  & \multicolumn{1}{c}{430}  \\
    
    \bottomrule
  \end{tabular}
\end{table}

\section{Metrics}

The evaluation of CoVoC includes both subjective and objective evaluations. All audio samples are evaluated using objective metrics, and a selected subset of 100 audio samples is used for subjective evaluations. During the objective evaluation, all audio samples are resampled to a uniform rate of 16 kHz.

\subsection{Objective Metrics}
The objective evaluation consisted of two aspects: pronunciation accuracy and timbre similarity. We employ Character Error Rate (CER) and cosine similarity for evaluation: 
\begin{itemize}
\item \textbf{Character Error Rate}: CER is computed between the ground truth transcript and the recognized transcript. We use an open-source paraformer-large model~\cite{paraformer} to recognize the synthesized speech into the corresponding transcription.
\item \textbf{Speaker Similarity (SIM)}: We employ the Resemblyzer tool~\cite{secs,generalized} to extract speaker embedding and compute the cosine similarity between the reference speech and generated speech.
\end{itemize}

\subsection{Subjective Metrics}
\label{sec:sub_eval}

For subjective evaluation, we conducted mean opinion score (MOS) tests to assess speech in four aspects: speech quality, speech naturalness, speaker similarity, and speech spontaneous style. 

\begin{itemize}
\item \textbf{Speech Naturalness (SN)}: 
Evaluate the naturalness of the generated speech. Consider whether the pronunciation is correct, if there is any ambiguity, if there are tone changes, and if the pauses sound natural. Assign a score from 1 (completely unnatural) to 5 (highly natural).

\item \textbf{Speech Quality (SQ)}: 
Evaluate the quality of the generated speech.
Determine if there is any electronic distortion and if the voice is clear.
Assign a score from 1 (very poor) to 5 (excellent).
\item \textbf{Speaker Similarity (SS)}: 
Evaluate the similarity between the generated speech and the target speaker, focusing on timbre and speaking style. As the speaker's audio contains only one sentence, the style may be somewhat uniform, so it is less crucial to focus on style similarity. Assign a score from 1 (not similar at all) to 5 (highly similar).
\item \textbf{Speech Spontaneous Style (SSS)}: 
Evaluate the colloquial characteristics of the speech. Consider whether the pronunciation of colloquial words such as "um," "uh," and "ah" sounds natural, whether laughter and non-rhythmic pauses are normal, and if the stress and rhythm resemble those of a real person. Assign a score from 1 (completely unnatural) to 5 (highly natural).
\end{itemize}

We invite 10 professional raters to listen to the generated samples and give a score for each audio sample. The subjective listening tests used the original audio submissions from the competition teams without any additional processing.
The final score (FS) shown in Table 1 is computed by a weighted sum of the 4 MOS scores across four aspects:
\begin{equation}
  \text{FS} = 0.25 \times \text{SN} + 0.25 \times \text{SQ} + 0.25 \times \text{SS} + 0.25 \times \text{SSS}
  \label{eq1}
\end{equation}

\section{Submitted systems}
A total of 11 teams submitted final results: 5 in the constrained track and 7 in the unconstrained track. These teams consisted of 7 from industry and 4 from academia. Detailed information about each team's system is presented in Table~\ref{tab:total_submint}.

As outlined in Section~\ref{sec:task}, the training data for the constrained track systems was limited to specified datasets, although open-source models were permitted. Only two systems reported training dataset sizes ranging from 10,000 to 300,000 hours in the unconstrained track.

Regarding acoustic models, most teams employed autoregressive (AR) text-to-semantics models, with only three opting for non-autoregressive (NAR) structures. Various methods were used at the waveform generation stage, including Codec~\cite{encodec} decoders, VITS~\cite{vits} decoders, and Flow Matching~\cite{flowmatching}, with no significant differences observed in objective and subjective evaluation outcomes. Each team's final score was determined by a weighted sum of four scores from subjective evaluations as detailed in Section~\ref{sec:sub_eval}.

Although not considered in the ranking, the challenge organizer also submitted a system with the highest score in the final subjective listening test. This system utilized a TorToise-like model framework and employed Single-Codec~\cite{single-codec} for speech tokenization. It featured frequency band expansion in the vocoder to improve audio quality and incorporated DSPGAN~\cite{dspgan} to upscale a 16k mel-spectrogram to a 48k high-fidelity waveform. In the final inference stage, in-context learning (ICL) was employed, prepending the target speaker's text and audio to the input text, thus treating speech synthesis as a continuation task.

\section{Results and Analysis}

\begin{figure}[h]
  \centering
  \includegraphics[width=\linewidth]{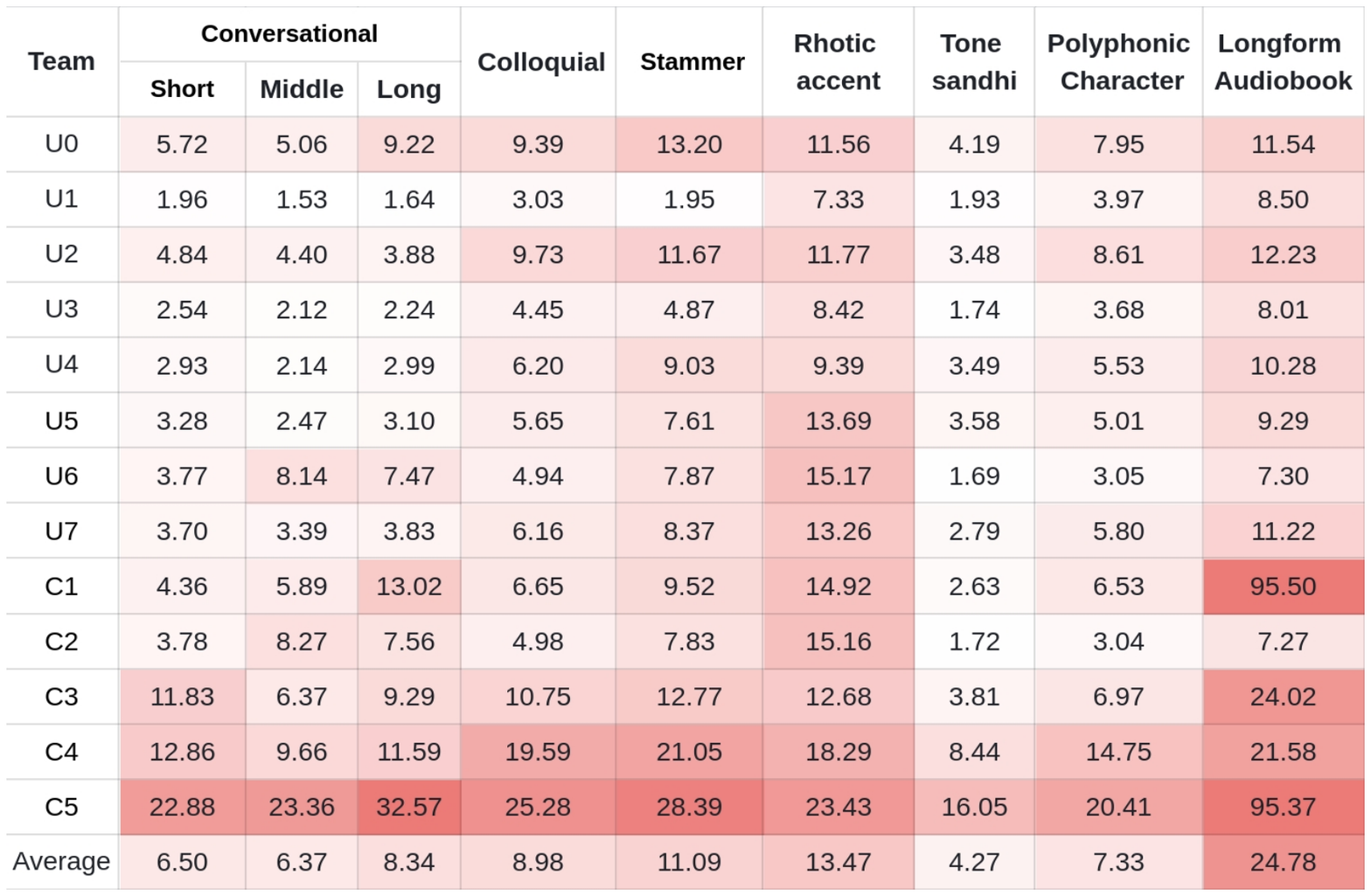}
  \caption{Average CER for each team across different text sets. The redder the color, the higher the CER.}
  \label{fig:CER_of_each_team}
\end{figure}

\begin{figure}[h]
  \centering
  \includegraphics[width=1.1\linewidth]{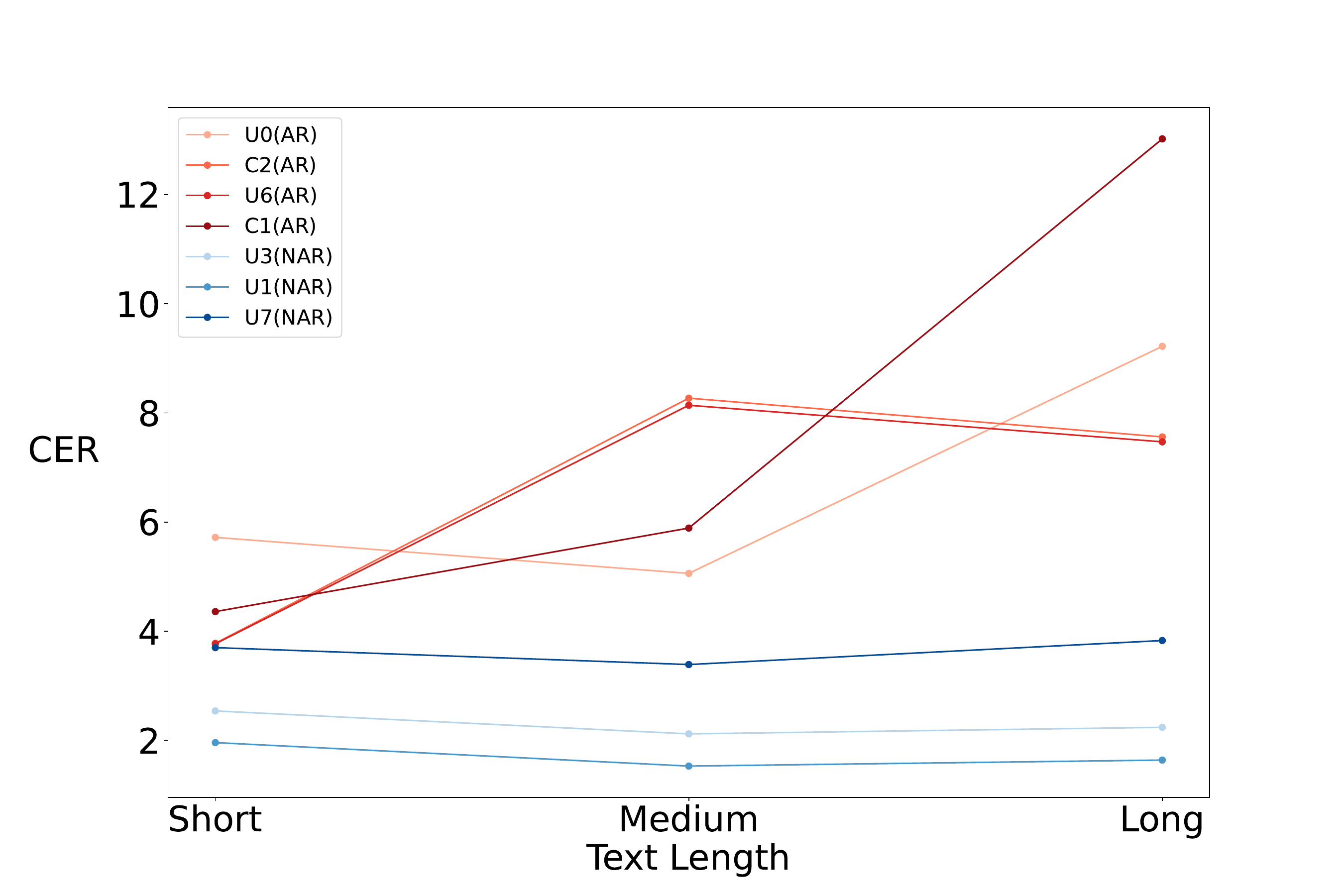}
  \caption{The CER results between different text lengths.}
  \label{fig:CERvstxt_len}
\end{figure}

We conducted a systematic comparative analysis of the submission results from all teams.
In terms of objective metrics, we analyzed the relationship between CER and text type, as well as the relationship between SIM and the length of the target speaker's audio.
In terms of subjective metrics, we found that the spontaneous style of speech is closely related to the speaking style of the target speaker.

\subsubsection{Analysis between CER and Text Type}

We calculated the average CER of each team on different types of test texts. The results are shown in Figure~\ref{fig:CER_of_each_team}. All participating teams performed well on the tone sandhi test set, while all teams showed an increased CER on the rhotic accent test set. This indicates that current systems learned well for the tone sandhi patterns with a large amount of paired training data, but their performance is relatively limited on the less frequent rhotic accent data.

Additionally, teams \textit{C1-we are NPC}, \textit{C3-THU-HCSI}, \textit{C4-ViveTTS}, and \textit{C5-SMIIP TTS} showed a significant increase in CER on the long-form audiobook test set, partly due to the small proportion of audiobook text types in the constrained track training and partly due to unresolved instability issues in AR models.

For the long texts, we found that the spectrogram is quite clear in the intervals between two consecutive sentences where there should be a pause in some submissions. We argue that these teams segmented the long text into smaller sentences for synthesis and then concatenated the short segments as the final results for submission. Despite this, CER still showed a significant positive correlation with text length.

We plotted the curve of the CER with text length across some typical AR and NAR systems as shown in Figure~\ref{fig:CERvstxt_len}. It can be seen that three NAR teams (U1-MASTER-TTS, U3-Orion, U7-hySoundClone) in the unconstrained track obtained low CER and were less affected by text length. 

\subsubsection{Analysis between SIM and Prompt Duration}
\begin{figure}[h]
  \centering
  \includegraphics[width=1.1\linewidth]{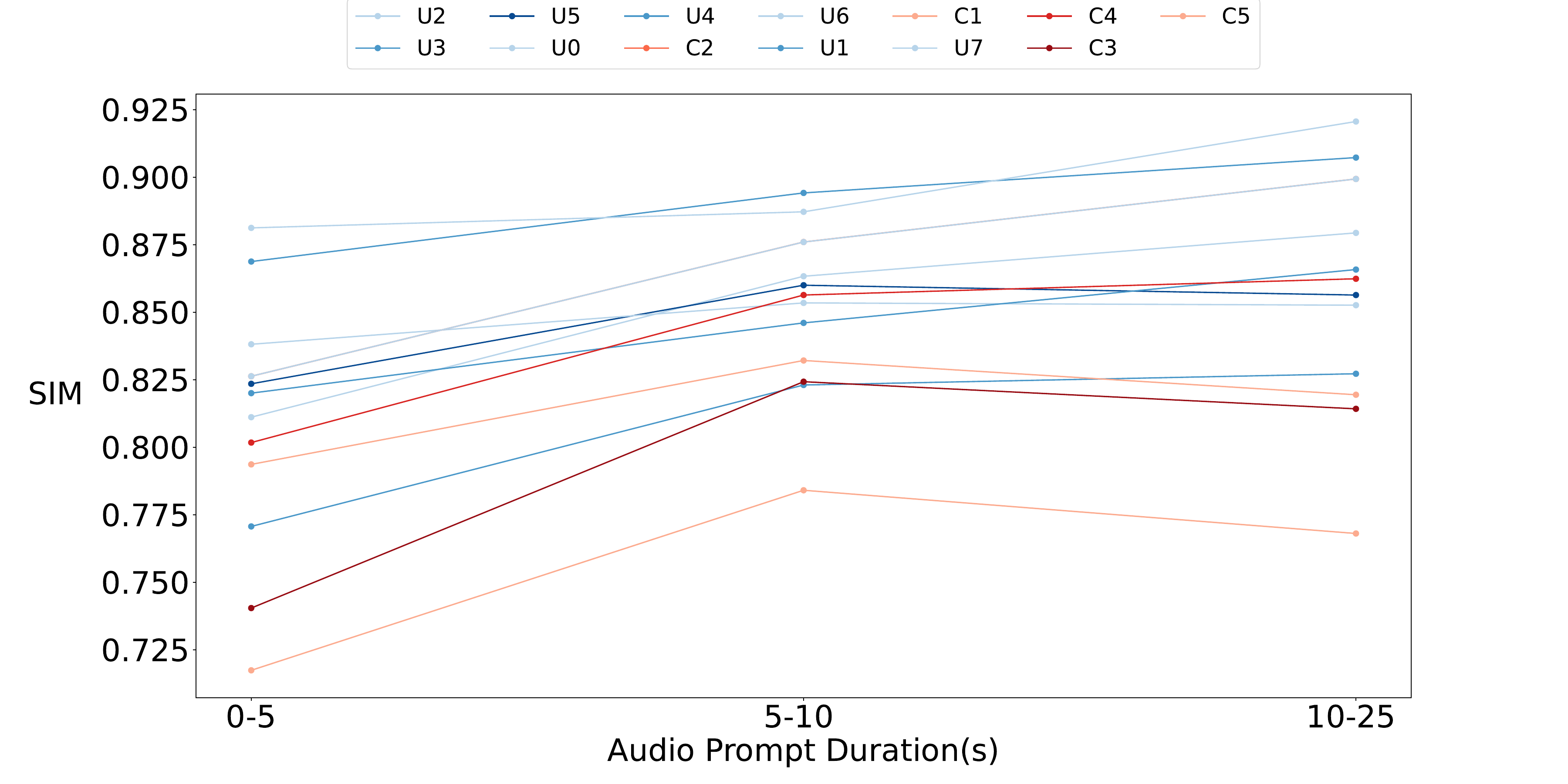}
  \caption{The similarity results between different duration lengths of the audio prompt.}
  \label{fig:team_sim_tend}
    
\end{figure}

In terms of the objective metric of speaker similarity, we found that as the duration of the target speaker's audio increases, the speaker similarity (SIM) tends to rise, as shown in Figure~\ref{fig:team_sim_tend}.
This trend is particularly noticeable when the target speaker's audio duration increases from the range of 0 to 5 seconds to the range of 5 to 10 seconds.

\subsubsection{Analysis between SSS and Prompt Style}
The MOS scores in 4 dimensions are listed in Table~\ref{tab:sub_evl_res}. As we can see, the scores of SN and SQ are generally higher than those of SS and SSS across all participating teams, indicating the naturalness and quality of the current speech synthesis systems have achieved significant advancements.

\begin{table}[h]
  \caption{Subjective evaluation results of the submitted systems.}
  \label{tab:sub_evl_res}
  \centering
  \begin{tabular}{ c c c c c c }
    \toprule
    \multicolumn{1}{c}{\textbf{Rank}}  & \multicolumn{1}{c}{\textbf{SN↑}}  & \multicolumn{1}{c}{\textbf{SQ↑}} & \multicolumn{1}{c}{\textbf{SS↑}} & \multicolumn{1}{c}{\textbf{SSS↑}} \\
    \midrule
    \multicolumn{1}{c}{U0}   & \multicolumn{1}{c}{ \textbf{4.07 ±0.10}} & \multicolumn{1}{c}{ 4.16 ±0.13} & \multicolumn{1}{c}{ 3.62 ±0.10} & \multicolumn{1}{c}{ \textbf{3.78 ±0.10}}  \\
    \midrule
    \multicolumn{1}{c}{U1}  & \multicolumn{1}{c}{ 3.84 ±0.11} & \multicolumn{1}{c}{ \textbf{4.23 ±0.15}} & \multicolumn{1}{c}{ 3.68 ±0.10} & \multicolumn{1}{c}{ 3.58 ±0.11}  \\
    \multicolumn{1}{c}{U2}  & \multicolumn{1}{c}{ \textbf{3.86 ±0.10}} & \multicolumn{1}{c}{ 4.11 ±0.14} & \multicolumn{1}{c}{ 3.47 ±0.11} & \multicolumn{1}{c}{ 3.64 ±0.10} \\
    \multicolumn{1}{c}{U3}  & \multicolumn{1}{c}{ 3.61 ±0.11} & \multicolumn{1}{c}{ 4.11 ±0.14} & \multicolumn{1}{c}{ 3.65 ±0.10} & \multicolumn{1}{c}{ 3.63 ±0.08}  \\
    \multicolumn{1}{c}{U4}  & \multicolumn{1}{c}{ 3.92 ±0.10} & \multicolumn{1}{c}{ 3.96 ±0.15} & \multicolumn{1}{c}{ 3.41 ±0.12} & \multicolumn{1}{c}{ \textbf{3.71 ±0.10}} \\
    \multicolumn{1}{c}{U5}  & \multicolumn{1}{c}{ 3.74 ±0.10} & \multicolumn{1}{c}{ 3.92 ±0.15} & \multicolumn{1}{c}{ 3.68 ±0.09} & \multicolumn{1}{c}{ 3.55 ±0.09}  \\
    \multicolumn{1}{c}{C1}  & \multicolumn{1}{c}{ 3.80 ±0.10} & \multicolumn{1}{c}{ 3.97 ±0.15} & \multicolumn{1}{c}{ 3.42 ±0.11} & \multicolumn{1}{c}{ 3.65 ±0.09}  \\
    \multicolumn{1}{c}{U6}  & \multicolumn{1}{c}{ 3.73 ±0.12} & \multicolumn{1}{c}{ 3.69 ±0.19} & \multicolumn{1}{c}{ 3.69 ±0.11} & \multicolumn{1}{c}{ 3.49 ±0.10}  \\
    \multicolumn{1}{c}{C2}  & \multicolumn{1}{c}{ 3.70 ±0.12} & \multicolumn{1}{c}{ 3.65 ±0.19} & \multicolumn{1}{c}{ \textbf{3.70 ±0.10}} & \multicolumn{1}{c}{ 3.48 ±0.10}  \\
    \multicolumn{1}{c}{C3}  & \multicolumn{1}{c}{ 3.80 ±0.11} & \multicolumn{1}{c}{ 3.84 ±0.16} & \multicolumn{1}{c}{ 3.49 ±0.12} & \multicolumn{1}{c}{ 3.33 ±0.12}  \\
    \multicolumn{1}{c}{U7}  & \multicolumn{1}{c}{ 3.37 ±0.12} & \multicolumn{1}{c}{ 3.80 ±0.14} & \multicolumn{1}{c}{ 3.54 ±0.11} & \multicolumn{1}{c}{ 3.36 ±0.10}  \\
    \multicolumn{1}{c}{C4}  & \multicolumn{1}{c}{ 3.27 ±0.12} & \multicolumn{1}{c}{ 3.52 ±0.17} & \multicolumn{1}{c}{ 3.04 ±0.13} & \multicolumn{1}{c}{ 3.25 ±0.10}  \\
    \multicolumn{1}{c}{C5}  & \multicolumn{1}{c}{ 3.41 ±0.14} & \multicolumn{1}{c}{ 3.35 ±0.19} & \multicolumn{1}{c}{ 2.88 ±0.14} & \multicolumn{1}{c}{ 2.95 ±0.13}  \\
    \bottomrule
  \end{tabular}
\end{table}

\begin{figure}[h]
  \centering
  \includegraphics[width=1\linewidth]{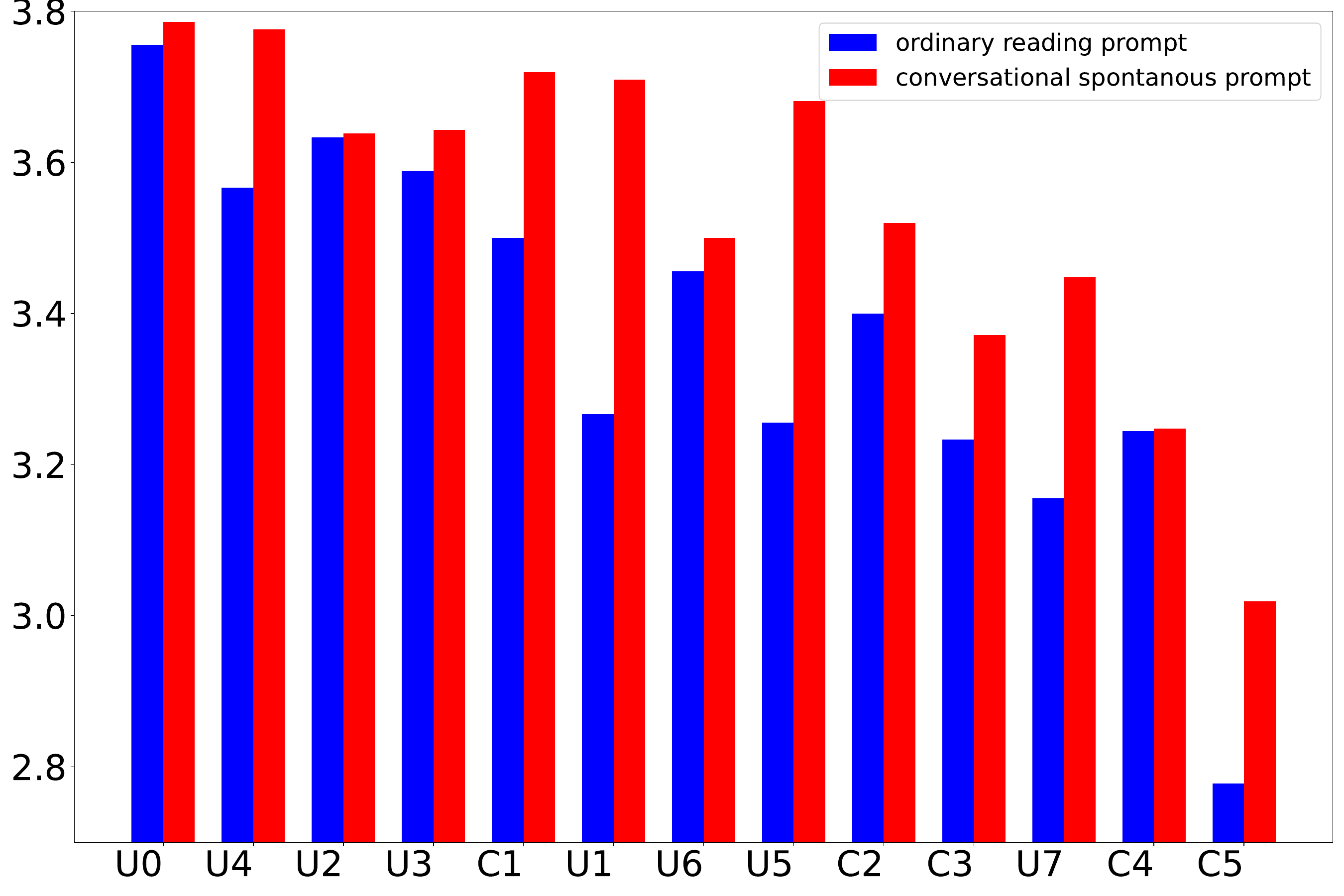}
  \caption{MOS score of SSS between two types of audio prompt.}
  \label{fig:sss_in_2}
\end{figure}

Comparing the speech spontaneous style (SSS) scores between two types of target speakers, i.e., the speakers with ordinary audio prompts and dialog prompts, we found that the scores for conversational spontanous style target speakers were consistently higher than those for ordinary reading style target speakers, as shown in Figure \ref{fig:sss_in_2}. 


In addition, we find that some teams, such as U1, U5, and U7, suffer significant degradation when the prompt is an ordinary speaking style. This indicates that the synthesized results of these systems strongly depend on prompt speech and exhibit poor stability. 
Furthermore, the two top-final-scoring autoregressive teams (U2, U4) perform  well in the spontaneous style, suggesting that autoregressive models still have some advantages in modeling specific speaking styles.


\section{Conclusion}

The CoVoC Challenge has successfully established a comprehensive benchmark for evaluating zero-shot voice cloning in spontaneous conversational contexts. Participation from various teams across academia and industry has highlighted significant advancements in generating high-quality, spontaneous-style speech. The release of standardized datasets, such as WenetSpeech4TTS and HQ-Conversations, has facilitated consistent training and evaluation, addressing a critical gap in the field of TTS research. The challenge's results highlight the effectiveness of both autoregressive and non-autoregressive models in different aspects of speech synthesis. Objective evaluations revealed that while non-autoregressive models generally achieve lower character error rates, autoregressive models often excel in maintaining speech spontaneous style over varying prompt styles. 
Through detailed analysis, we identified key areas for improvement, such as handling long texts and rare speech patterns like rhotic accents. The strong performance of non-autoregressive models on text length indicates potential pathways for future research. In summary, the CoVoC Challenge has established a robust foundation for future studies in zero-shot TTS, promoting the development of models capable of generating natural, spontaneous speech. Continued efforts in dataset expansion, model innovation, and evaluation methodologies will further enhance the capabilities of conversational voice cloning systems.

\bibliographystyle{IEEEtran}

\bibliography{mybib}

\begin{thebibliography}{10}
\providecommand{\url}[1]{#1}
\csname url@samestyle\endcsname
\providecommand{\newblock}{\relax}
\providecommand{\bibinfo}[2]{#2}
\providecommand{\BIBentrySTDinterwordspacing}{\spaceskip=0pt\relax}
\providecommand{\BIBentryALTinterwordstretchfactor}{4}
\providecommand{\BIBentryALTinterwordspacing}{\spaceskip=\fontdimen2\font plus
\BIBentryALTinterwordstretchfactor\fontdimen3\font minus \fontdimen4\font\relax}
\providecommand{\BIBforeignlanguage}[2]{{%
\expandafter\ifx\csname l@#1\endcsname\relax
\typeout{** WARNING: IEEEtran.bst: No hyphenation pattern has been}%
\typeout{** loaded for the language `#1'. Using the pattern for}%
\typeout{** the default language instead.}%
\else
\language=\csname l@#1\endcsname
\fi
#2}}
\providecommand{\BIBdecl}{\relax}
\BIBdecl

\bibitem{tacotron}
Y.~Wang, R.~J. Skerry{-}Ryan, D.~Stanton, Y.~Wu, R.~J. Weiss, N.~Jaitly, Z.~Yang, Y.~Xiao, Z.~Chen, S.~Bengio, Q.~V. Le, Y.~Agiomyrgiannakis, R.~Clark, and R.~A. Saurous, ``Tacotron: Towards end-to-end speech synthesis,'' in \emph{Proc. Interspeech}, 2017, pp. 4006--4010.

\bibitem{fs}
Y.~Ren, Y.~Ruan, X.~Tan, T.~Qin, S.~Zhao, Z.~Zhao, and T.~Liu, ``Fastspeech: Fast, robust and controllable text to speech,'' in \emph{Proc. NeurIPS}, 2019, pp. 3165--3174.

\bibitem{vits}
J.~Kim, J.~Kong, and J.~Son, ``Conditional variational autoencoder with adversarial learning for end-to-end text-to-speech,'' in \emph{Proc. {ICML}}, 2021, pp. 5530--5540.

\bibitem{tortoise}
J.~Betker, ``Better speech synthesis through scaling,'' \emph{arXiv preprint arXiv:2305.07243}, 2023.

\bibitem{valle}
C.~Wang, S.~Chen, Y.~Wu, Z.~Zhang, L.~Zhou, S.~Liu, Z.~Chen, Y.~Liu, H.~Wang, J.~Li, L.~He, S.~Zhao, and F.~Wei, ``Neural codec language models are zero-shot text to speech synthesizers,'' \emph{arXiv preprint arXiv:2301.02111}, 2023.

\bibitem{spontts}
H.~Li, X.~Zhu, L.~Xue, Y.~Song, Y.~Chen, and L.~Xie, ``Spontts: modeling and transferring spontaneous style for tts,'' in \emph{ICASSP 2024-2024 IEEE International Conference on Acoustics, Speech and Signal Processing (ICASSP)}.\hskip 1em plus 0.5em minus 0.4em\relax IEEE, 2024, pp. 12\,171--12\,175.

\bibitem{sponlm}
W.~Li, P.~Yang, Y.~Zhong, Y.~Zhou, Z.~Wang, Z.~Wu, X.~Wu, and H.~Meng, ``Spontaneous style text-to-speech synthesis with controllable spontaneous behaviors based on language models,'' \emph{arXiv preprint arXiv:2407.13509}, 2024.

\bibitem{congjian_conversation}
J.~Cong, S.~Yang, N.~Hu, G.~Li, L.~Xie, and D.~Su, ``Controllable context-aware conversational speech synthesis,'' in \emph{Interspeech}.\hskip 1em plus 0.5em minus 0.4em\relax {ISCA}, 2021, pp. 4658--4662.

\bibitem{wenetspeech4tts}
L.~Ma, D.~Guo, K.~Song, Y.~Jiang, S.~Wang, L.~Xue, W.~Xu, H.~Zhao, B.~Zhang, and L.~Xie, ``Wenetspeech4tts: A 12,800-hour mandarin tts corpus for large speech generation model benchmark,'' \emph{arXiv preprint arXiv:2406.05763}, 2024.

\bibitem{wenetspeech}
B.~Zhang, H.~Lv, P.~Guo, Q.~Shao, C.~Yang, L.~Xie, X.~Xu, H.~Bu, X.~Chen, C.~Zeng, D.~Wu, and Z.~Peng, ``{WENETSPEECH:} {A} 10000+ hours multi-domain mandarin corpus for speech recognition,'' in \emph{{ICASSP}}.\hskip 1em plus 0.5em minus 0.4em\relax {IEEE}, 2022, pp. 6182--6186.

\bibitem{megatts}
Z.~Jiang, Y.~Ren, Z.~Ye, J.~Liu, C.~Zhang, Q.~Yang, S.~Ji, R.~Huang, C.~Wang, X.~Yin \emph{et~al.}, ``Mega-tts: Zero-shot text-to-speech at scale with intrinsic inductive bias,'' \emph{arXiv preprint arXiv:2306.03509}, 2023.

\bibitem{delightfultts}
Y.~Liu, Z.~Xu, G.~Wang, K.~Chen, B.~Li, X.~Tan, J.~Li, L.~He, and S.~Zhao, ``Delightfultts: The microsoft speech synthesis system for blizzard challenge 2021,'' \emph{arXiv preprint arXiv:2110.12612}, 2021.

\bibitem{dit}
W.~Peebles and S.~Xie, ``Scalable diffusion models with transformers,'' in \emph{{ICCV}}.\hskip 1em plus 0.5em minus 0.4em\relax {IEEE}, 2023, pp. 4172--4182.

\bibitem{musicgen}
J.~Copet, F.~Kreuk, I.~Gat, T.~Remez, D.~Kant, G.~Synnaeve, Y.~Adi, and A.~D{\'{e}}fossez, ``Simple and controllable music generation,'' in \emph{NeurIPS}, 2023.

\bibitem{speartts}
E.~Kharitonov, D.~Vincent, Z.~Borsos, R.~Marinier, S.~Girgin, O.~Pietquin, M.~Sharifi, M.~Tagliasacchi, and N.~Zeghidour, ``Speak, read and prompt: High-fidelity text-to-speech with minimal supervision,'' \emph{Trans. Assoc. Comput. Linguistics}, vol.~11, pp. 1703--1718, 2023.

\bibitem{flowmatching}
Y.~Lipman, R.~T.~Q. Chen, H.~Ben{-}Hamu, M.~Nickel, and M.~Le, ``Flow matching for generative modeling,'' in \emph{{ICLR}}.\hskip 1em plus 0.5em minus 0.4em\relax OpenReview.net, 2023.

\bibitem{lauragpt}
Q.~Chen, Y.~Chu, Z.~Gao, Z.~Li, K.~Hu, X.~Zhou, J.~Xu, Z.~Ma, W.~Wang, S.~Zheng \emph{et~al.}, ``Lauragpt: Listen, attend, understand, and regenerate audio with gpt,'' \emph{arXiv preprint arXiv:2310.04673}, 2023.

\bibitem{paraformer}
Z.~Gao, S.~Zhang, I.~McLoughlin, and Z.~Yan, ``Paraformer: Fast and accurate parallel transformer for non-autoregressive end-to-end speech recognition,'' \emph{arXiv preprint arXiv:2206.08317}, 2022.

\bibitem{secs}
Y.~Jia, Y.~Zhang, R.~J. Weiss, Q.~Wang, J.~Shen, F.~Ren, Z.~Chen, P.~Nguyen, R.~Pang, I.~L{\'{o}}pez{-}Moreno, and Y.~Wu, ``Transfer learning from speaker verification to multispeaker text-to-speech synthesis,'' in \emph{NeurIPS}, 2018, pp. 4485--4495.

\bibitem{generalized}
L.~Wan, Q.~Wang, A.~Papir, and I.~L. Moreno, ``Generalized end-to-end loss for speaker verification,'' in \emph{2018 IEEE International Conference on Acoustics, Speech and Signal Processing (ICASSP)}.\hskip 1em plus 0.5em minus 0.4em\relax IEEE, 2018, pp. 4879--4883.

\bibitem{encodec}
A.~D{\'{e}}fossez, J.~Copet, G.~Synnaeve, and Y.~Adi, ``High fidelity neural audio compression,'' \emph{Trans. Mach. Learn. Res.}, vol. 2023, 2023.

\bibitem{single-codec}
H.~Li, L.~Xue, H.~Guo, X.~Zhu, Y.~Lv, L.~Xie, Y.~Chen, H.~Yin, and Z.~Li, ``Single-codec: Single-codebook speech codec towards high-performance speech generation,'' \emph{arXiv preprint arXiv:2406.07422}, 2024.

\bibitem{dspgan}
K.~Song, Y.~Zhang, Y.~Lei, J.~Cong, H.~Li, L.~Xie, G.~He, and J.~Bai, ``{DSPGAN}: A gan-based universal vocoder for high-fidelity tts by time-frequency domain supervision from dsp,'' in \emph{Proc. {ICASSP}}, 2023.

\end{thebibliography}


\end{document}